# Ultrafast electrooptic dual-comb interferometry

Vicente Durán[1], Santiago Tainta[2], and Victor Torres-Company[1,*]

1. Department of Microtechnology and Nanoscience, Chalmers University of Technology, SE 41296 Gothenburg, Sweden
2. Department of Electrical and Electronic Engineering, Universidad Pública de Navarra, 31006 Pamplona, Spain

*torresv@chalmers.se

**Abstract**

The femtosecond laser frequency comb has enabled the 21st century revolution in optical synthesis and metrology. A particularly compelling technique that relies on the broadband coherence of two laser frequency combs is dual-comb interferometry. This method is rapidly advancing the field of optical spectroscopy and empowering new applications, from nonlinear microscopy to laser ranging. Up to now, most dual-comb interferometers were based on modelocked lasers, whose repetition rates have restricted the measurement speed to ~ kHz. Here we demonstrate a novel dual-comb interferometer that is based on electrooptic frequency comb technology and measures consecutive complex spectra at a record-high refresh rate of 25 MHz. These results pave the way for novel scientific and metrology applications of frequency comb generators beyond the realm of molecular spectroscopy, where the measurement of ultrabroadband waveforms is of paramount relevance.

**Introduction**

With the advent of the femtosecond laser frequency comb, scientists and engineers today have at their disposal a tool for optical synthesis and metrology with a level of performance that was previously achieved only by a few specialized laboratories in the world[1,2]. The spectrum of a laser frequency comb is formed by a set of evenly spaced narrow-linewidth frequencies that maintain the phase coherence across the bandwidth. There is an important attribute that distinguishes an optical frequency comb from any other multi-wavelength laser source, namely the possibility to tune independently the spacing between lines (or repetition rate) and the central optical frequency[3].

The dual-comb spectrometer (also known as dual-comb interferometer) is an instrumentation technique that unlocks the full potential of laser frequency combs for high-precision spectroscopy[4-7]. The features of a dual-comb spectrometer are simply unattainable with state-of-the-art Fourier transform spectrometers. For example, it offers a spectral resolution capable to resolve the individual frequency components of the comb. It can also scan terahertz bandwidth spectra at high signal to noise ratio in relatively fast acquisition times (from milliseconds to a few seconds) because it is free from mechanical moving parts[6]. This technique has been successfully applied at different wavelength regions[8-12] and in modern spectroscopy applications, such as remote sensing of greenhouse gases[13] and nonlinear hyperspectral microscopy[14]. An additional characteristic of a dual-comb spectrometer is that it can retrieve the spectral phase of the sample under test. This has triggered applications different than molecular spectroscopy, such as coherent LIDAR[15-17], vibrometry[18] or the analysis of optical telecommunication components[19].

Most dual-comb interferometers presented to date make use of fiber or Ti:Sa modelocked laser technology. The repetition rate frequency of these laser sources typically lies in the 10-100 MHz range. This comb spacing provides a spectral resolution more than adequate for molecular spectroscopy, but introduces a fundamental limit in the speed that it takes to capture the optical spectrum (typically in the millisecond range). Longer acquisition times are often necessary in order to improve the signal-to-noise ratio of the measurement by coherently averaging successive spectra[20]. Long acquisition times introduce stringent demands on the phase locking of the two combs employed in the dual-comb spectrometer, whose relative phase drifts need to be compensated for. This has been solved by using combs locked to external optical references (and therefore providing long-term stability[6]) or in free-running modelocked lasers by either applying real-time signal processing techniques[21] or adapting dynamically the sampling clock rate of the dual-comb spectrometer to the relative drift of the combs' offset and spacing[22]. Outstanding signal-to-noise ratios corresponding to > 20 bits have been reported for measurements accumulated in the course of 24 hrs using real-time signal processing[21].

The electrooptic modulation method is an alternative technique for producing coherent frequency combs[23-26]. This technique was introduced in the 70s and it basically consists of modulating a continuous-wave laser with a chain of electrooptic modulators that are driven by an external radio-frequency oscillator. The central frequency is given by the laser frequency and the radio-frequency oscillator provides the line spacing of the comb. This technique does not rely on modelocking and hence allows for tuning in an independent manner the offset and line spacing of the comb[27]. An electrooptic comb is assembled with standard optical telecommunication equipment and indeed has been used as a source for lightwave communications[28] and radio-frequency photonics[27,29]. This technique has gained momentum in the last years thanks to advances in fiber laser technology, high-performance lithium niobate modulators and microwave dielectric resonator oscillators[27]. The optical spectrum of an electrooptic comb can be substantially broadened in a highly nonlinear fiber[30,31] and there is recent progress towards achieving self-referencing[32], an important ingredient for absolute metrology. Several groups have recently implemented a dual-comb spectrometer with electrooptic frequency comb technology[19,33-35]. The rationale lies in the robustness and simplicity in the hardware implementation, and the fact that a single laser feeds the two combs, achieving the necessary phase locking by default[33] without the need for complex feedback stabilization mechanisms.

Here we uncover an additional benefit of an electrooptic dual-comb spectrometer, namely the possibility to operate at unprecedented speeds. This is due to the fact that an electrooptic comb generator operates at repetition rates in the 10 GHz range and therefore has a few lines covering a broad bandwidth. We show that this trade in spectral resolution can be used to increase the measurement speed while affecting neither the measurement bandwidth nor the accuracy. We report complex spectral measurements spanning over a terahertz bandwidth with record refresh rates in the nanosecond regime using standard radio-frequency equipment. These results pave the way to utilize the dual-comb technique in different scientific applications that require robust and accurate measurements of broadband waveforms at ultra-high speeds.

## Results

### *Dual-comb spectroscopy*

The operation principle of dual-comb spectroscopy can be understood in two equivalent ways[6]. In the time domain, the complex amplitude of a sample is encoded on the spectrum of a train of pulses with repetition rate $f_s$. This optical signal interferes with a reference train of pulses that has a slightly different repetition rate $f_r = f_s \pm \delta f$. The frequency offset, $\delta f$, makes the pulses from each comb to overlap on the detector at varying time delays. As a result, the interference signal becomes the electric-field cross-correlation between the sample and the reference combs[19]. A Fourier transform analysis of every interferogram provides the complex amplitude of the sample assuming the reference is known. Therefore, a dual-comb spectrometer essentially works as a 'virtual' scanning interferometer or, alternatively, as linear coherent sampling.

In the frequency-domain picture, the interference between the sample and reference leads to a multi-heterodyne detection process, since each line of the signal comb beats with the lines from the reference. The resulting beat notes are distributed in groups (Nyquist zones) along the radio-frequency region, leading to a downconversion of the optical frequencies. Usually, one only considers the part of the radio-frequency spectrum that corresponds to the first Nyquist zone (which spans from dc to half the smaller of the two comb repetition rates). In an electrooptic dual-comb spectrometer, the same laser feeds the two combs. Then the spectra share the central frequency and pairs of lines in the upper and lower sidebands produce beat notes at exactly the same frequency. To avoid this ambiguity, we follow the scheme first proposed in Ref. 19 and shift the laser's frequency with the aid of an acousto-optic modulator before going into one of the corresponding comb generators, as shown in Fig. 1. This shift moves the interference between the central comb lines away from dc and creates two interlaced radio-frequency combs in the downconversion process[12]. Each radio-frequency comb has a frequency spacing of $\delta f$, but a different offset that depends on the acousto-optic frequency $f_{AOM}$ as indicated in Fig. 1(a). Taking into account that the refresh rate will be given by the frequency location of the beat note closest to dc, we note that both $f_{AOM}$ and $\delta f$ must be carefully chosen to optimize the measurement rate. This is particularly important for electrooptic combs, which have a few number of lines and therefore can be widely spread across the detection bandwidth.

Our experimental setup is shown in Fig. 1(b). We set two electrooptic combs operating at ~25 GHz repetition rate with an offset of $\delta f = 100$ MHz and $f_{AOM} = 25$ MHz. In this manner, it is the frequency offset between the central lines of the combs that determines the maximum refresh rate. Each comb provides 55 lines within -10 dB, covering a bandwidth of ~1.4 THz. A high power (1W) narrow linewidth laser feeds the electrooptic combs[26] and ensures shot-noise operation even after the sample comb passes through the sample. An important aspect of the setup here presented is that it is self-synchronized and the sampling stage is commensurate to the frequency offsets. This alleviates the digital signal processing and any drift in the repetition rate of the comb is automatically taken into account at the sampling stage.

### *Optical arbitrary waveform characterization*

The spectroscopy examples in this work correspond to synthetic waveforms programmed in the sample arm with the aid of a reconfigurable complex filter (a pulse shaper). In our case the shaper resolution is better than the comb spacing. This corresponds to the line-by-line regime and the synthesized waveforms may have a duty cycle approaching 100%[36]. We choose this coherent regime because it introduces great challenges for any measurement technique owing to the possible overlap at the period boundaries between consecutive pulses in the train[37-40].

Figure 2(a) shows an oscilloscope trace measured during 10 μs for a train of pulses with a cubic spectral phase programmed in a line-by-line manner. The periodicity of the registered electrical signal is evident in the lower inset. The duration of an interferogram is $T = 1/f_{AOM} = 40$ ns, so a complete temporal trace recorded by the oscilloscope contains 5000 interferograms, each one formed by four consecutive 10-ns electrical waveforms. Figure 2(b) shows the temporal intensity profiles for a 20-ps window calculated from 2500 interferograms, which are processed individually. The signal coming from the pulse shaper is also recorded by a commercial optical sampling scope. The result provided by the dual-comb technique has significantly better stability and temporal resolution (~ 600 fs) because of the tight phase locking between combs and the inherent broadband operation.

Next we explore the synthesis of optical waveforms with duty cycles of ~ 100%. We codify onto the pulse shaper a spectral phase profile corresponding to a sinusoidal phase function with an abrupt phase change of π, shown in the blue curve in Fig. 3(a). For each signal comb line, we retrieve a set of 2500 phases, whose mean values are shown in Fig. 3(a) as red points. In the course of these measurements, we realized that the phase sensitivity of the dual-comb technique is better than the nominal phase setting accuracy of the pulse shaper (~0.1 rad). To estimate the precision of our measurements, we calculated the standard deviation of the retrieved phases for each comb line. The result can be observed in Fig. 3(b). It is apparent that the phase error increases as one moves away from the central frequency. The mean value of the phase error, averaged over all comb lines, is 50 mrad, which is equivalent to an optical path difference of ~12 nm at 1545 nm (or ~$\lambda$/125). This indicates the potential of this technique to realize different phase-sensitive precision measurements.

Figure 3(c) compares an autocorrelation measurement of the generated waveform (red line) with the autocorrelation function theoretically calculated from the retrieved spectral phase (blue line) from the dual-comb technique. A commercial autocorrelator, placed at the end of the signal arm, measures a curve composed of 1000 points during 50 s. The theoretical line, in its turn, is calculated from the spectral phases obtained for a 100-µs temporal trace. The nonzero values of the autocorrelation function at the edges of the period demonstrate interference between neighbor pulses, i.e. a duty cycle close to 100%. The close match between the experimental and calculated curves corroborates the accuracy of our method when compared to a well-established pulse characterization technique. Figure 3(d) shows the intensity pulse profile of 2500 waveforms calculated from the dual-comb technique. The fine line of this multivalued curve indicates very low intensity fluctuations (below 0.01 at the highest peak).

*Sensitivity analysis*

One advantage of performing fast measurements is the possibility of using coherent averaging to increase the signal to noise ratio (at the expense of reducing the effective measurement refresh rate). This tradeoff between sensitivity and speed is an inherent feature of dual-comb spectroscopy. The ultrafast single-shot acquisition speeds achieved in our method permit to average thousands of waveforms, hence recover extremely weak signals (orders of magnitude lower than the local oscillator) and still operate at an effective refresh rate in the kHz regime. This improvement in sensitivity is especially desirable for illuminating targets at low power levels (an essential issue, for instance, to avoid damage in biological specimens) or when highly absorbing or scattering samples are considered.

To analyze the performance of our system, we systematically reduce the light power $P_s$ coming from the pulse shaper with the aid of an attenuator, as indicated in Fig. 1(b). The spectral phase chosen for this experiment is a continuous sinusoidal function programmed in a line-by-line manner. The sensitivity can be enhanced by means of a preamplified detection scheme so that the signal's power level corresponds to pulses containing on average just a few photons (see Methods). Figure 4 shows the phase accuracy achieved by this configuration when the signal power is progressively decreased. At an effective refresh rate of 1 MHz, the mean phase error remains below 0.1 rad up to -39 dBm. However, for weaker signals, the amplified spontaneous emission noise in the preamplifier becomes predominant and the phase error increases. By averaging more spectral phases the signal to noise ratio improves and therefore the error decreases. At an effective refresh rate of 100 kHz, the mean phase error exceeds the pulse shaper accuracy for $P_s \lesssim$ -48 dBm. By reducing the effective refresh rate to 25 kHz (i.e. four times more waveforms averaged), it is still possible to recover the programmed phase function at -50 dBm, that is, at pulse waveform containing 3 photons on average. The recovered spectral phase for this minimum signal power is shown in the inset in Fig. 4. The difference between the experimental phase values and those programmed onto the pulse shaper is comparable to that observed, for example, in Fig. 3(a) for -10 dBm.

**Discussion**

Any dual-comb spectrometer shows a fundamental tradeoff between number of lines, optical bandwidth and speed. Hitherto most configurations have been designed to provide a spectral resolution in the MHz regime covering terahertz bandwidth spectra, at the expense of the refresh rate (in the millisecond regime). We have demonstrated that the electrooptic dual-comb technique allows measuring broadband optical waveforms with similar accuracy, but the limited number of lines (~ 50-100) can be used as leverage to boost the refresh rate by several orders of magnitude. To put things in context, the resulting spectral resolution, bandwidth and refresh rate here reported are comparable to the performance achieved by the dispersive Fourier transformation technique[41]. A key distinctive aspect in dual-comb spectroscopy is that by default the setup is sensitive to the spectral phase of the sample and does not require distributed amplification to operate in the shot-noise regime. In addition, with lithium niobate modulators, the wavelength operation range can be designed anywhere within ~ 0.8- 2 μm.

The fastest technique that measures optical waveforms in a line-by-line manner is full-field coherent arbitrary waveform measurement[40]. This technique is highly suitable for coherent communication applications where the signal waveform needs to be measured at the baud rate. The hardware implementation is challenging though, since it requires N tightly synchronized coherent receivers with a bandwidth equal to the comb rate for N comb lines. In contrast, the dual-comb technique is multi-heterodyne and therefore a single, relatively low-frequency-bandwidth acquisition unit (in this report 3 GHz) is required to measure a broadband waveform composed of tens of lines.

The scaling of the refresh rate in the dual-comb technique is highly favorable and recent progress in fiber parametric comb generators[42] (whose repetition rates operate beyond the limit of current electrooptic modulators) hint that, for comb containing tens of lines, it should be possible to bring the measurement rate to the sub-nanosecond regime utilizing a state-of-the-art sampling unit with 40 GHz analog bandwidth. Microresonator frequency combs[43] offer the prospect of combining photonic integration with ultra-high-repetition rates, and depending on the material they can operate in different regions of the electromagnetic spectrum[44].

Regarding applications, the ability to measure arbitrary terahertz bandwidth waveforms in the sub-microsecond regime opens up new prospects in ultrafast metrology. We mention for example, frequency domain reflectometry and optical coherence tomography; the measurement of telecommunication equipment in absence of slow environmental drifts affecting the measurement; dynamic profilometry of surfaces in industrial machining; and high-speed phase-sensitive imaging.

## Methods

### Electrooptic frequency comb generators

Each comb generator is composed of an intensity modulator followed by a pair of phase modulators. The commercially available devices are based on lithium niobate electrooptic modulators and are specially designed to handle high power (both from the microwave source and the input laser). They are all driven by an external clock at ~ 25 GHz. The clock is a commercial low-phase-noise dielectric resonator oscillator. The continuous-wave laser is a low RIN, low linewidth (~10 kHz), laser centered at 1545 nm whose power is boosted to 1W by an erbium-doped amplifier. The intensity modulator is biased to provide a train of pseudo-square pulses with ~50% of duty cycle. The chirp of the phase modulators is aligned to the square pulses with tunable microwave phase shifters. The intensity modulator blocks the light when the chirp from the phase modulators is mostly linear. In this way, the comb spectrum becomes relatively flat[25]. The use of two phase modulators in tandem enables to increase the effective modulation index and hence the comb bandwidth. Our arrangement provides 55 lines at -10dB bandwidth (or 1.4 THz optical bandwidth). The losses of the combs are ~ -16 dB and the optical carrier-to-noise ratio per line is higher than 60 dB (measurement limited by the dynamic range of the optical spectrum analyzer).

### Dual-comb interferometer

The frequency offset between the combs is $\delta f = 100$ MHz. An acousto-optic modulator is also included in the signal arm to shift the laser frequency. The sample considered in our experiments is a commercial pulse shaper (Finisar 4000 S) with 10 GHz resolution and 1 GHz accuracy, which is used to synthesize the spectrum of the signal comb in a line-by-line manner. The light emerging from the pulse shaper is controlled by a variable optical attenuator and combined with the reference comb on a 50:50 coupler. The interference is measured by a balanced detector (BD, from u2t Photonics AG, model BPDV2150R). The balanced detection avoids the 3-dB power penalty inherent to the mixing process and eliminates the unwanted dc term that comes with the interference signal. In both arms, polarization controllers are inserted at the entrance of each comb and just before the final coupler to optimize the optical power. The radio-frequency signal generated by the detector is digitized by an 8 bits oscilloscope with 3 GHz bandwidth (LeCroy Wavemaster 8300), which registers a temporal trace during a total record time of 200 μs at a sampling rate of 10 GS/s. When we use the detector in a preamplified configuration, an erbium-doped fiber amplifer is inserted in the signal arm. We ensure that the average power arriving to the balanced detector remains constant (around 0 dBm). Depending on the degree of signal attenuation, an extra fiber amplifier can be included in cascade. This combination of preamplification and balanced detection is especially beneficial, since only the beating between the amplified spontaneous emission and the local oscillator gives a relevant noise term[45]. The output of the photodetector is amplified by a stage of three low-noise microwave amplifiers to optimize the number of bits in the detection unit.

In the radio-frequency domain [yellow box in Fig. 1(b)], the offset in repetition rate frequencies is extracted via a mixer. One half of the signal at the output of the mixer is used as both external clock and trigger for the oscilloscope. The other half passes through a frequency divider for producing a radio-frequency signal whose frequency is reduced by a factor 4, which ensures that the frequency shift introduced by the acousto-optic modulator is commensurate with the repetition rate offset between the combs.

**System calibration**

Our dual-comb spectrometer requires a calibration process to work. To this end, an oscilloscope trace is recorded when the pulse shaper acts as an all-pass filter. This measurement gives the relative complex amplitude between the two interferometer arms. Depending on the system configuration, the pulse shaper acting on the signal, and another one incorporated in the reference arm [not shown in Fig. 1(b)], can be used to shape each comb spectrum. This enables to compensate in a convenient manner during the course of different experiments for the dispersion unbalance between the arms of the interferometer. Afterwards, superimposed to the calibration profile implemented on the signal pulse shaper, we add the spectral function under study. For example, for the results shown in Fig. 2(b), the programmed phase profile is the superposition of two terms. The first one is an approximately quadratic function for producing a train of transformed-limited pulses. The other term is the cubic phase imparted onto de spectrum, which is plotted in the left inset.

**Data processing**

The process that enables to recover the signal programmed onto the pulse shaper can be briefly summarized as follows. We use one half of the trace measured by the oscilloscope, which contains 2500 interferograms, each one composed of four consecutive 10-ns waveforms. For each interferogram, the spectral complex amplitude is recovered through an FFT. In order to extract the successive interferograms, we use the 100-MHz clock signal generated by the mixer shown in Fig. 1(b). Conventional tools in FFT analysis, as zero-padding or phase unwrapping algorithms, are employed to obtain the results presented in Fig. 3. When the recorded traces are very noisy, as in the sensitivity experiments presented in Fig. 4, a preprocessing of the measured data becomes necessary. In that case, we calculate the FFT of the complete temporal trace and remove undesired spurious RF frequencies. The filtering process is accomplished by means of a bandpass comb filter, which is composed of teeth that are centered on the frequencies of interest. The bandpass is the same for every teeth and its bandwidth will limit the number of waveforms that can be averaged. Hence this filtering is especially beneficial at effective refresh rates below a few MHz.

## Acknowledgements

The authors are thankful to Peter Andrekson for technical discussions and support.

This work has been funded by the Swedish Research Council (VR, grant number 048701). VD acknowledges funding from a Marie Curie Intra European Fellowship (PIEF-GA-2013-625121) and VTC a Marie Curie Career Integration grant (PCIG13-GA-2013-618285). ST acknowledges funding from the Spanish Ministry of Science and Innovation (project TEC2010-21303-04-01).

## References

[1] T. W. Hänsch. Nobel lecture: Passion for precision. *Rev. Mod. Phys.* 78, 1297-1309 (2006).

[2] J. L. Hall. Nobel lecture: Defining and measuring optical frequencies. *Rev. Mod. Phys.* 78, 1279-1295 (2006).

[3] N. R. Newbury. Searching for applications with a fine-tooth comb. *Nature Photon.* 5, 186-188 (2011).

[4] S. Schiller. Spectrometry with frequency combs. *Opt. Lett.* 27, 766–768, (2002).

[5] F. Keilmann, C. Gohle, and R. Holzwarth. Time-domain mid-infrared frequency-comb spectrometer. *Opt. Lett.* 29, 1542–1544 (2004).

[6] I. Coddington, W. C. Swann, and N. R. Newbury. Coherent multiheterodyne spectroscopy using stabilized optical frequency combs. *Phys. Rev. Lett.* 100, 013902 (2008).

[7] B. Bernhardt, A. Ozawa, P. Jacquet, M. Jacquey, Y. Kobayashi, T. Udem, R. Holzwarth, G. Guelachvili, T. W. Hänsch, and N. Picqué. Cavity-enhanced dual-comb spectroscopy. *Nature Photon.* 4, 55–57 (2010).

[8] A. Schliesser, M. Brehm, F. Keilmann, and D. van der Weide. Frequency-comb infrared spectrometer for rapid, remote chemical sensing. *Opt. Express* 13, 9029–9038 (2005).

[9] F. Zhu, T. Mohamed, J. Strohaber, A. A. Kolomenskii, T. Udem, and H. A. Schuessler. Real-time dual frequency comb spectroscopy in the near infrared. *Appl. Phys. Lett.* 102, 121116 (2013).

[10] S. Potvin and J. Genest. Dual-comb spectroscopy using frequency-doubled combs around 775 nm. *Opt. Express* 21, 30707-30715 (2013).

[11] Z. Zhang, T. Gardiner, and D. T. Reid. Mid-infrared dual-comb spectroscopy with an optical parametric oscillator. *Opt. Lett.* 38, 3148–3150 (2013).

[12] G. Villares, A. Hugi, S. Blaser, and J. Faist. Dual-comb spectroscopy based on quantum-cascade-laser frequency combs. *Nature Commun.* 5:5192 (2014).

[13] G. B. Rieker, F. R. Giorgetta, W. C. Swann, J. Kofler, A. M. Zolot, L. C. Sinclair, E. Baumann, C. Cromer, G. Petron, C. Sweeny, P. P. Tans, I. Coddington, and N. R. Newbury. Frequency-comb-based remote sensing of greenhouse gases over kilometer air paths. *Optica* 1, 290-298 (2014).

[14] T. Ideguchi, S. Holzner, B. Bernhardt, G. Guelachvili, N. Picqué, and T. W. Hänsch. Coherent Raman spectro-imaging with laser frequency combs. *Nature* 502, 355–358 (2013).

[15] I. Coddington, W. C. Swann, L. Nenadovic, and N. R. Newbury. Rapid and precise absolute distance measurements at long range. *Nature Photon.* 3, 351–356 (2009).

[16] S. Boudreau, S. Levasseur, C. Perilla, S. Roy, and J. Genest. Chemical detection with hyperspectral lidar using dual frequency combs. *Opt. Express* 21, 7411–7418 (2013).

[17] T.-A. Liu, N. R. Newbury, and I. Coddington. Sub-micron absolute distance measurements in sub-millisecond times with dual free-running femtosecond Er fiber-lasers. *Opt. Express* 19, 18501–18509 (2011).

[18] S. Boudreau and J. Genest. Range-resolved vibrometry using a frequency comb in the OSCAT configuration. *Opt Express* 22, 8101–8113 (2014).

[19] F. Ferdous, D. E. Leaird, C.-B. Huang, and a M. Weiner. Dual-comb electric-field cross-correlation technique for optical arbitrary waveform characterization. *Opt. Lett.* 34, 3875–3877 (2009).

[20] I. Coddington, W. Swann, and N. Newbury. Coherent dual-comb spectroscopy at high signal-to-noise ratio. *Phys. Rev. A* 82, 043817 (2010).

[21] J. Roy, J.-D. Deschênes, S. Potvin, and J. Genest. Continuous real-time correction and averaging for frequency comb interferometry. *Opt. Express* 20, 21932-21939 (2012).

[22] T. Ideguchi, A. Poisson, G. Guelachvili, N. Picqué, and T. W. Hänsch. Adaptive real-time dual-comb spectroscopy. *Nature Commun.* 5:3375 (2014).

[23] T. Kobayashi, H. Yao, K. Amano, Y. Fukushima, A. Morimoto, and T. Sueta. Optical pulse compression using high-frequency electro-optic phase modulation. *IEEE J. Quantum Electron.* 24, 382-387 (1988).

[24] T. Yamamoto, T. Komukai, K. Takada, and A. Suzuki. Spectrally flattened phase-locked multi-carrier light generator with phase modulators and chirped fibre Bragg grating. *Electron. Lett.* 43, 1040-1042 (2007).

[25] T. Otsuji, M. Yaita, T. Nagatsuma, and E. Sano. 10-80 Gb/s highly extinctive electro-optic pulse pattern generator. *IEEE J. Sel. Top. In Quantum Electron.* 2, 643-649 (1996).

[26] A. J. Metcalf, V. Torres-company, D. E. Leaird, S. Member, A. M. Weiner. High-power broadly tunable electrooptic frequency comb generator. *IEEE J. Sel. Top. In Quantum Electron.* 19, 350036 (2013).

[27] V. Torres-Company and A. M. Weiner. Optical frequency comb technology for ultra-broadband radio-frequency photonics. *Laser & Photon. Rev.* 8, 368–393 (2014).

[28] T. Ohara, H. Takara, T. Yamamoto, H. Masuda, T. Morioka, M. Abe, and H. Takahashi. Over-1000-channel ultradense WDM transmission with supercontinuum multicarrier source. *J. Lightwave Technol.* 24, 2311-2317 (2006).

[29] E. Hamidi, D. E. Leaird, and A. M. Weiner. Tunable programmable microwave photonic filters based on an optical frequency comb. *IEEE Trans. Microwave Theory Techn.* 58, 3269-3278 (2010).

[30] A. Ishizawa, T. Nishikawa, A. Mizutori, H. Takara, H. Nakano, T. Sogawa, A. Takada, and M. Koga. Generation of 120-fs laser pulses at 1-GHz repetition rate derived from continous wave laser diode. *Opt. Express* 19, 22402-22409 (2011).

[31] R. Wu, V. Torres-Company, D. E. Leaird, and A. M. Weiner. Supercontinuum-based 10-GHz flat-topped optical frequency comb generator. *Opt. Express* 21, 6045-6052 (2013).

[32] K. Beha, D. C. Cole, F. N. Baynes, P. Del'Haye, A. Rolland, T. M. Fortier, F. Quinlan, S. A. Diddams, and S. B. Papp. Towards self-referencing a 10-GHz electro-optic comb. ED-1a.3 Mon *Proc. CLEO Europe* (2015).

[33] D. A. Long, A. J. Fleisher, K. O. Douglass, S. E. Maxwell, K. Bielska, J. T. Hodges, and D. F. Plusquellic. Multiheterodyne spectroscopy with optical frequency combs generated from a continuous-wave laser. *Opt. Lett.* 39, 2688–2690 (2014).

[34] P. Martin-Mateos, M. Ruiz-Llata, J. Posada-Roman, and P. Acedo. Dual-comb architecture for fast spectroscopic measurements and spectral characterization. *IEEE Photonics Technol. Lett.* 27, 1309–1312 (2015).

[35] G. Millot, S. Pitois, M. Yan, T. Hovannysyan, A. Bendahmane, T. W. Hänsch, and N. Picqué. Frequency-agile dual-comb spectroscopy. *Arxiv 1505.07213* (2015).

[36] Z. Jiang, C.-B. Huang, D. E. Leaird, and A. M. Weiner. Optical arbitrary waveform processing of more than 100 spectral comb lines. *Nature Photon.* 1, 463-467 (2007).

[37] V. R. Supradeepa, D. E. Leaird, and A. M. Weiner. Single shot amplitude and phase characterization of optical arbitrary waveforms. *Opt. Express* 17, 14434-14443 (2009).

[38] N. K. Fontaine, R. P. Scott, J. P. Heritage, and S. J. B. Yoo. Near quantum-limited, single-shot coherent arbitrary optical waveform measurements. *Opt. Express* 17, 12332-12344 (2009).

[39] V. R. Supradeepa, C. M. Long, D. E. Leaird, and A. M. Weiner. Self-referenced characterization of optical frequency combs and arbitrary waveforms using a simple, linear, zero-delay implementation of spectral shearing interferometry. *Opt. Express* 18, 18171-18179 (2010).

[40] N. K. Fontaine, R. P Scott, L. Zhou, F. M. Soares, J. P. Heritage, and S. J. B. Yoo. Real-time full-field arbitrary optical waveform measurement. *Nature Photon.* 4, 248-254 (2010).

[41] D. R. Solli, J. Chou, and B. Jalali. Amplified wavelength-time transformation for real-time spectroscopy. *Nature Photon.* 2, 48-51 (2007).

[42] V. Ataie, E. Myslivets, B. P. -P. Kuo, N. Alic, and S. Radic. Spectrally equalized frequency comb generation in multistage parametric mixer with nonlinear pulse shaping. *J. Lightwave Technol.* 32, 840-846 (2014).

[43] T. J. Kippenberg, R. Holzwarth, and S. A. Diddams. Microresonator-based optical frequency combs. *Science* 332, 555-559 (2011).

[44] C. Y. Wang, T. Herr, P. Del'Haye, A. Schliesser, J. Hofer, R. Holzwarth, T. W. Hänsch, N. Picqué, and T. J. Kippenberg. Mid-infrared optical frequency combs at 2.5 μm based on crystalline microresonators. *Nature Commun.* 4:1345 (2013).

[45] S. Yamashita and T. Okoshi. Suppression of Beat Noise from Optical Amplifiers Using Coherent Receivers. *J. Lightwave Technol.* 12, 1029–1035 (1994).

**Figure 1**

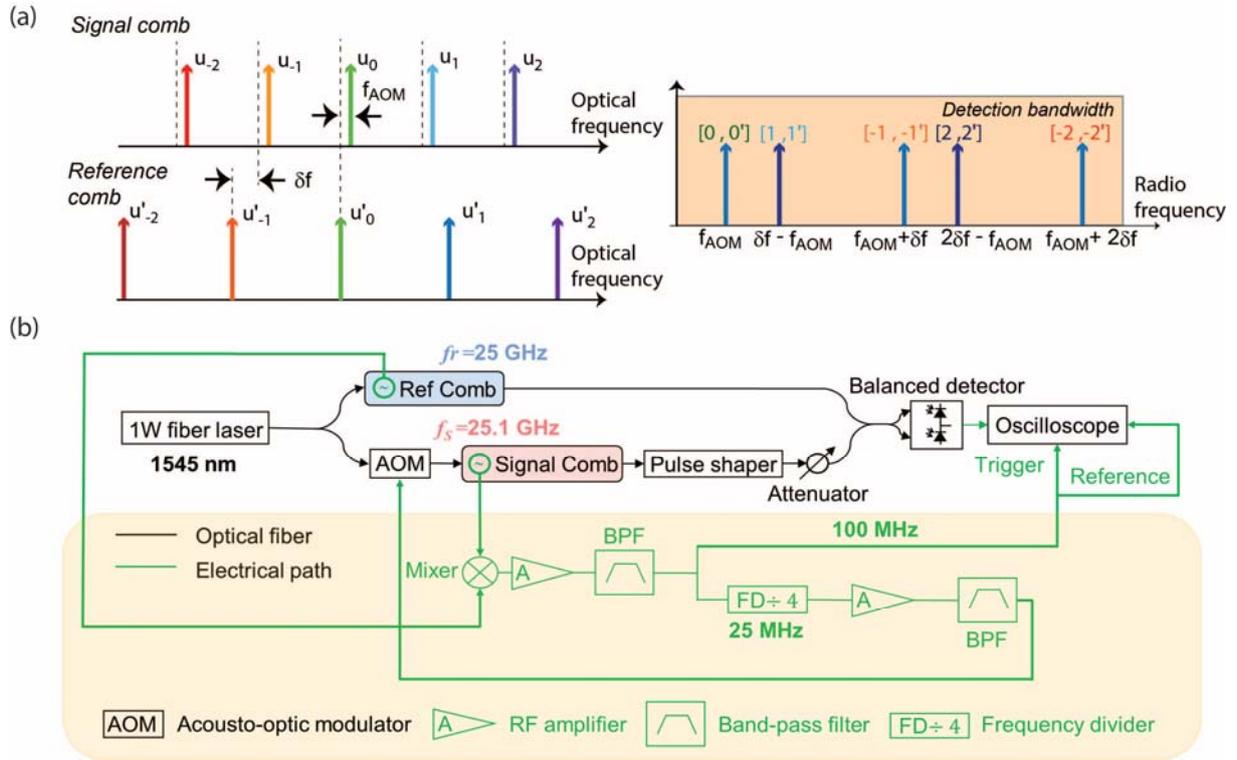

Figure 1. (a) In the frequency domain, the multiheterodyne interference of the two combs is interpreted as a downconversion of optical frequencies. The offset $\delta f$ between the repetition rates of the two combs generates a set of beat notes in the radio-frequency region. In our approach, both combs are fed by the same laser. In such a case, one of the two comb spectra must be shifted by an amount $f_{AOM}$ in order to ensure unambiguous downconversion. The result is two interlaced combs in the radio-frequency domain. The radio frequencies are labeled by pairs of numbers that identify the beat between the mother optical comb lines $u_i$ and $u'_i$ ($i=$-2,…,2). (b) Schematic of the experimental setup. Basically, it consists of a fiber heterodyne interferometer, being the signal and the local oscillator two electro-optic frequency combs fed by a high-power continuous-wave laser. The spectroscopy sample is an optical pulse shaper that allows us to synthesize the optical lines of the signal comb with a spectral resolution compatible with the repetition rate of the comb. The radio-frequency circuit is designed to be self-synchronized with the sampling frequencies commensurate to the repetition rates of the comb, hence helping the digital signal processing stage.

**Figure 2**

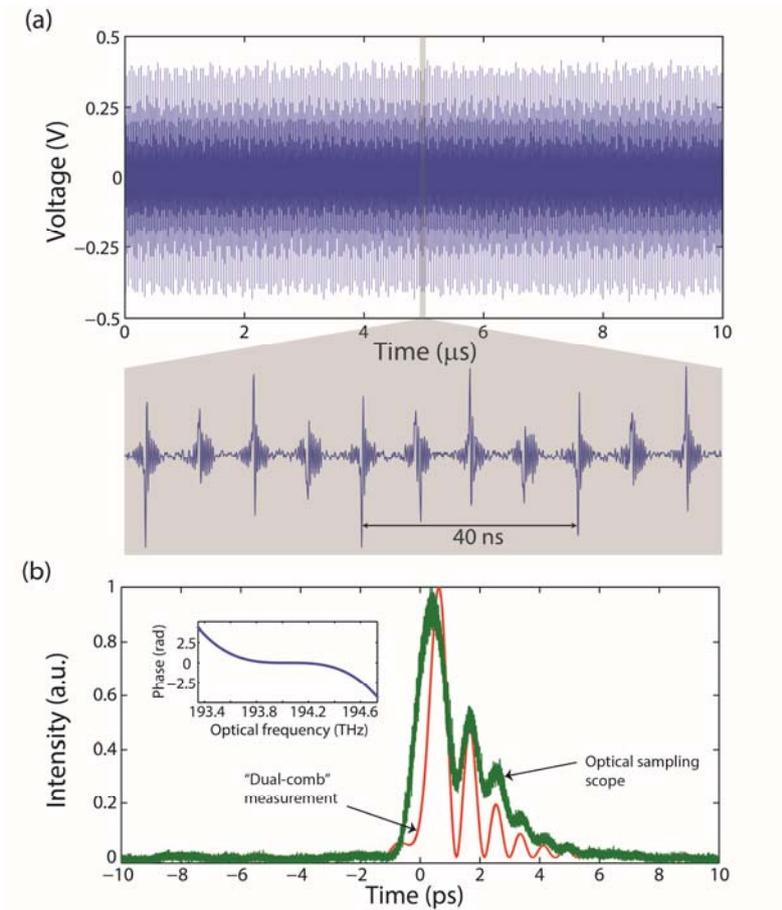

Figure 2. Example of an arbitrary optical waveform characterization of a train of pulses with cubic phase. (a) Voltage measured by the oscilloscope during 10 μs (5% of the total record length). The periodicity of the recorded signal can be observed in the zoomed image included in the lower inset. (b) Intensity profile inside a 20-ps time window corresponding to the programmed spectral phase. The red profile is formed by the set of curves calculated from the recovered spectral phases corresponding to 2500 individual waveforms. The programmed phase profile is shown on the left. The green curve is the temporal signal recorded by a commercial optical sampling scope with limited temporal resolution (1 ps).

**Figure 3**

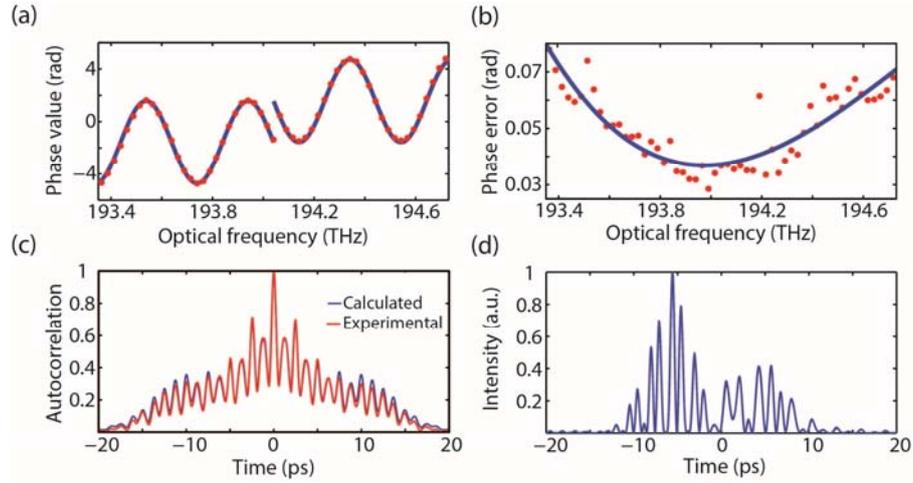

Figure 3. Single-shot characterization of a 100% duty factor waveform. (a) Phase profile imparted onto the signal spectrum by the pulse shaper (blue curve) and spectral phases obtained by averaging the values recovered from 2500 individually processed waveforms (red points). (b) Standard deviation of the recovered phase for each comb line. (c) Autocorrelation function corresponding to the waveform generated by the line-by-line shaping. (d) 2500 overlaid intensity profiles built from the electrooptic dual-comb measurements. In this measurement, the light power coming from the reference comb is $P_r$ = 1 mW, while the sample power is one order of magnitude smaller, $P_s$ = 100 µW.

**Figure 4**

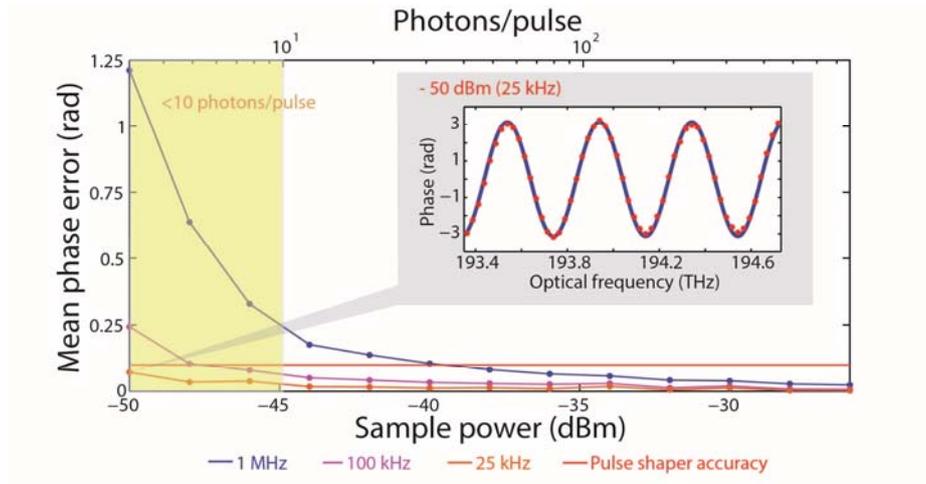

Fig. 4. Analysis of the system sensitivity. The plot shows the mean phase error as a function of the signal power for three different effective refresh rates. The shadowed zone corresponds to signal pulses that on average have only a few photons. The recovered phase for $P_s$ = -50 dBm (at 25 kHz effective refresh rate) is included in the inset. In this graph, the match of the phase values with the programmed curve is lower than the pulse shaper accuracy (0.1 rad) along the bandwidth of the comb.